\documentclass[preprint,aps,nofootinbib]{revtex4}
\usepackage{graphicx}
\usepackage{color}

\def\bold#1{\setbox0=\hbox{$#1$}%
     \kern-.025em\copy0\kern- -\wd0
     \kern.05em\copy0\kern-\wd0
     \kern-.025em\raise.0433em\box0 }

\def\tanb{\tan\beta}
\def\bea{\begin{eqnarray}}  
\def\eea{\end{eqnarray}}
\newcommand{\ba}{\begin{array}}
\newcommand{\ea}{\end{array}}

\def\noi{\noindent}
\def\what{\widehat}
\def\wtil{\widetilde}
\def\eps{\epsilon}
\def\vev#1{\langle #1 \rangle}
\def\beq{\begin{equation}}
\def\eeq{\end{equation}}
\def\bit{\begin{itemize}}
\def\eit{\end{itemize}}
\def\ben{\begin{enumerate}}
\def\een{\end{enumerate}}
\def\sign{{\rm sign}}
\def\lam{\lambda}
\def\kap{\kappa}
\def\cnone{\widetilde \chi_1^0}

\def\mcnone{m_{\cnone}}

\def\what{\widehat}

\def\anti{\overline}
\def\br{BR}
\def\gam{\gamma}

\def\vev#1{\langle #1 \rangle}

\newcommand{\gev}{\,\mbox{GeV}}

\def\ga{\mathrel{\raise.3ex\hbox{$>$\kern-.75em\lower1ex\hbox{$\sim$}}}}
\def\la{\mathrel{\raise.3ex\hbox{$<$\kern-.75em\lower1ex\hbox{$\sim$}}}}

\def\m12{m_{1\!/2}}

\def\lsim{\mathrel{\raise.3ex\hbox{$<$\kern-.75em\lower1ex\hbox{$\sim$}}}}
\def\gsim{\mathrel{\raise.3ex\hbox{$>$\kern-.75em\lower1ex\hbox{$\sim$}}}}

\begin{document}
\begin{titlepage}
\pagestyle{empty}
\baselineskip=21pt
\rightline{hep-ph/0509024}
\rightline{UCD--2005--06}
\rightline{FERMILAB-PUB-05-350-A}
\vskip 1in
\begin{center}
{\large{\bf  \boldmath
Light Neutralino Dark Matter in the NMSSM
}}
\end{center}
\begin{center}
\vskip 0.2in 
{\bf  John F. Gunion$^1$, Dan Hooper$^2$, Bob McElrath$^1$}
\vskip 0.1in 
$^1${\it Department of Physics, University of California,
Davis, CA 95616, USA}\\
$^2${\it Department of Physics, Oxford University, Oxford OX1-3RH, UK; Particle
Astrophysics Center, Fermi National Accelerator Laboratory, Batavia, IL
60510-0500, USA}

\vskip 0.2in {\bf Abstract}
\end{center}
\baselineskip=18pt

Neutralino dark matter is generally assumed to be relatively heavy, with
a mass near the electroweak scale. This does not necessarily need to be
the case, however. In the Next-to-Minimal Supersymmetric Standard Model
(NMSSM) and other supersymmetric models with an extended Higgs sector, a very light CP-odd Higgs
boson can naturally arise making it possible for a very light neutralino
to annihilate efficiently enough to avoid being overproduced in the
early Universe.  

In this article, we explore the characteristics of a supersymmetric
model needed to include a very light neutralino, 100 MeV $< \mcnone
<$ 20 GeV, using the NMSSM as a prototype. We discuss the most important
constraints from Upsilon decays, $b \rightarrow s \gamma$, $B_s \rightarrow \mu^+ \mu^-$ and the magnetic moment of the muon, and find that a light bino or singlino neutralino is allowed, and can be generated with the appropriate relic density.

It has previously been shown that the positive detection of dark matter claimed by the DAMA collaboration can be reconciled with other direct dark matter experiments such as CDMS II if the dark matter particle is rather light, between about 6 and 9 GeV. A singlino or bino-like neutralino could easily fall within this range of masses within the NMSSM. Additionally, models with sub-GeV neutralinos may be
capable of generating the 511 keV gamma-ray emission observed from the
galactic bulge by the INTEGRAL/SPI experiment.

We also point out measurements which can be performed immediately at CLEO, BaBar
and Belle using existing data to discover or significantly constrain this
scenario.

\end{titlepage}
\baselineskip=18pt
\setcounter{footnote}{0}

\section{Introduction}

Despite the substantial effort which has gone into its detection,
the nature of dark matter remains unknown \cite{review}. The dark matter
candidates which have received the most attention fall into the category
of Weakly Interacting Massive Particles (WIMPs), which can emerge from a
variety of theoretical frameworks, including supersymmetry. Of the
supersymmetric candidates for dark matter, the lightest neutralino is
often considered to be the most attractive.

Neutralinos produced in the early Universe must annihilate into Standard Model particles
at a sufficient rate to avoid overproducing the density of dark matter. Within the framework of the Minimal Supersymmetric Standard Model (MSSM), the lightest
neutralino can annihilate through a variety of channels, exchanging
other sparticles, $Z$ bosons, or Higgs bosons. The masses of sparticles
such as sleptons or squarks, as well as the masses of Higgs bosons, are
limited by collider constraints, with typical lower limits of around
$\sim$100 GeV. For lighter neutralinos, it becomes increasingly
difficult for these heavy propagators to generate neutralino
annihilation cross sections that are large enough. The most efficient
annihilation channel for very light neutralinos in the MSSM is the
$s$-channel exchange of a pseudoscalar Higgs boson. It has been shown that
this channel can, in principle, be sufficiently efficient to allow
for neutralinos as light as 6 GeV \cite{lightmssmdm}. Such models
require a careful matching of a number of independent parameters,
however, making viable models with neutralinos lighter than $\sim$20 GeV
rather unlikely \cite{Hooper:2002nq}. Measurements of
rare $B$-decays are also particularly constraining in this regime.  If we
do not require that the LSP be the dominant component of dark matter,
its mass can be zero~\cite{Gogoladze:2002xp}.

More generally speaking, Lee and Weinberg have demonstrated that a
fermionic dark matter candidate which annihilates through its couplings
to the weak gauge bosons must be heavier than a few GeV to avoid
over-closing the Universe~\cite{2gevfdm}. Therefore, if a neutralino is
to be very light, it requires another annihilation channel which enables
it to sufficiently annihilate in the early Universe. This can be
provided within the context of the Next-to-Minimal Supersymmetric
Standard Model (NMSSM) by the lightest of the two CP-odd Higgs
bosons, which can be considerably lighter than 
the single CP-odd Higgs boson of the MSSM without
violating collider constraints. Furthermore, it has been shown that
models within the NMSSM which require the smallest degree of fine tuning
often contain a light CP-odd Higgs boson \cite{finetuning}.

In addition to these theoretical arguments, there are experimental
motivations to consider light dark matter particles. The observation of
511 keV gamma rays from the galactic bulge~\cite{Jean:2003ci} indicates
the presence of a Gaussian profile of low-velocity positrons throughout
our galaxy's inner kiloparsec. It is challenging to explain this
observation with traditional astrophysics \cite{511astro}.  Annihilating
\cite{511annihilate} or decaying \cite{511decay} dark matter particles
have been suggested as a possible source of these positrons. If such a
dark matter particle were in the mass range usually considered, however,
their annihilation would produce positrons with far too much energy to
annihilate at rest. Furthermore, they would almost certainly generate
far too many gamma-rays and violate the constraints placed by EGRET
\cite{Beacom:2004pe}. Thus a dark matter candidate capable of generating
the observed 511 keV line must be exceptionally light.

Additionally, it has been shown that the claims of dark matter detection
made by the DAMA collaboration \cite{dama} can be reconciled with null
results of CDMS II \cite{cdms} and other experiments if one considers a
WIMP lighter than approximately 10 GeV \cite{gondolosd,gondolo}. 

In this article, we explore the phenomenology of supersymmetric models
with a neutralino in the mass range of 100 MeV to 20 GeV within the
context of the NMSSM. We find that many such models can be found which
are not highly fine tuned and are consistent with all constraints
including direct collider searches, rare decays, and relic abundance
considerations. We find examples of consistent models in which a light
neutralino can potentially produce the 511 keV emission observed by INTEGRAL as well
as models that can potentially reconcile DAMA with CDMS II. 
However,
we have not found models in which all these observations can
be simultaneously explained.

\section{Neutralino Dark Matter with a Singlet Higgs}

The simplest possible extension of the particle content of the MSSM is the
addition of a new gauge singlet chiral supermultiplet.  There are several ways
to do this including the NMSSM~\cite{nmssmus,nmssm}, the MNSSM (Minimal
Non-minimal Supersymmetric Standard Model)~\cite{mnssm} and larger
models~\cite{MSSM4S} with interesting implications for dark
matter.~\cite{Barger:2005hb} For concreteness and the availability of dominant
1-loop and 2-loop corrections to the higgs sector via the code
NMHDECAY,~\cite{Ellwanger:2004xm} we choose to study the NMSSM.  

Adding a Higgs singlet is attractive for
several reasons. Most interesting, perhaps, is that it provides an
elegant solution to the $\mu$-problem present in the MSSM
\cite{muproblem}. Additionally, the ``little fine tuning problem'', which
results in the MSSM from the lack of a detection of a CP-even Higgs at
LEP II, is less severe within the NMSSM \cite{finetuning,littletuning}, and is
completely absent if the lightest CP-odd Higgs is light enough to allow $H \rightarrow
A_1 A_1$ decays~\cite{finetuning}. Thirdly, baryogenesis considerations leave the MSSM in
disfavor, requiring the right handed stop squark to be lighter than the
top quark and the Higgs lighter than about 117 GeV \cite{mssmbary}.
Recent studies of baryogenesis within the NMSSM indicate that parameter
points with a light singlet Higgs and a corresponding light neutralino
are favored \cite{Funakubo:2005pu}.  Finally, the domain wall problem~\cite{domainwalls}
in the NMSSM can be avoided by the introduction of appropriate
non-renormalizable Planck-suppressed operators, and imposing a discrete
R-symmetry on them.~\cite{Panagiotakopoulos:1998yw}
  
In the NMSSM, the physical CP-even and CP-odd Higgs states are mixtures
of MSSM-like Higgses and singlets. The lightest neutralino therefore
has, in addition to the four MSSM components, a singlino component which is the 
superpartner of the singlet Higgs. The eigenvector of the lightest 
neutralino, $\cnone$, in terms of gauge eigenstates is:
\begin{equation} 
    \cnone = \epsilon_u \tilde{H}^0_u + \epsilon_d
    \tilde{H}^0_d + \epsilon_W \tilde{W}^0   + \epsilon_B \tilde{B} +
    \epsilon_s \tilde{S}, 
\end{equation}
where $\epsilon_u$, $\epsilon_d$ are the up-type and down-type higgsino
components, $\epsilon_W$, $\epsilon_B$ are the wino and bino components
and $\epsilon_s$ is the singlet component of the lightest neutralino.


Likewise, for the lightest CP-even Higgs state we can define:
\begin{equation} 
    H_1 = \left[\xi_u \Re \left({ H_u^0 \over \sqrt 2 } - v_u\right) + \xi_d
    \Re \left({
    H_d^0 \over \sqrt 2} - v_d\right) + \xi_s \Re \left({S \over \sqrt
    2} - x\right)\right].
\end{equation}
Here, $\Re$ denotes the real component of the respective state, and we
take vacuum expectation values to be those of the complex states (e.g.
$v = \sqrt{v_u^2+v_d^2} \simeq 174$ GeV). 

Lastly, we can write the lightest CP-odd Higgs as:
\begin{equation}
\label{as}
A_1 = \cos \theta_A A_{\rm MSSM} + \sin \theta_A A_s, 
\end{equation}
where $A_s$ is the CP-odd piece of the singlet and $A_{\rm MSSM}$ is the state that would be
the MSSM pseudoscalar Higgs if the singlet were not present. $\theta_A$ is the
mixing angle between these two states.  There is also a third imaginary linear
combination of $H_u^0$, $H_d^0$ and $S$ that we have removed by a
rotation in $\beta$.  This field becomes the longitudinal component of
the $Z$ after electroweak symmetry is broken.

The NMSSM can contain either an
approximate global $U(1)$ R-symmetry in the limit that the Higgs-sector
trilinear soft SUSY breaking terms are small, or a $U(1)$ Peccei-Quinn
symmetry in the limit that the cubic singlet term in the superpotential
vanishes \cite{nmssmaxion}.  In either case, one ends up with the lightest
CP-odd Higgs boson, $A_1$, as the pseudo-goldstone boson of this broken
symmetry, which can be very light.  In some regions of the NMSSM parameter
space, one can also get the lightest CP-even state, $H_1$, to be very
light as well. This is discussed in more detail in
Sec.~\ref{seclighta}.
As shown in Sec.~\ref{seclightchi}, it is easy to get a light largely
singlino LSP in the $U(1)_{PQ}$ symmetry limit. 

While we confine our analysis to the NMSSM, it should be noted that such
symmetries are generically present in other singlet models such as the
Minimal Non-minimal Supersymmetric Standard Model (MNSSM) \cite{mnssm}.
The combination of a light $A_1$ and a light neutralino is not uncommon
in a wide class of models with extra singlets and/or extra gauge groups
\cite{MSSM4S}. 
Implications of such models for the relic neutralino
density have been considered in~\cite{Barger:2004bz}.

The NMSSM is defined by the superpotential
\beq
\lambda \widehat H_u \widehat H_d \widehat S+{\kappa\over 3} \widehat
S^3
\label{superv}
\eeq
and associated soft-supersymmetry-breaking terms
\beq
\lambda A_\lambda H_u H_d S +{\kappa\over 3}A_\kappa S^3,
\eeq
where the hatted objects are chiral superfields and unhatted objects
are their scalar components. An effective $\mu$ parameter
(as defined by the superpotential form $\mu \what H_u \what H_d$
of the MSSM) 
is generated from the first term
of Eq.~(\ref{superv}) when $\vev{S}\equiv x$ is non-zero: $\mu=\lam x$.
We follow the sign conventions for NMSSM parameters of
Refs.~\cite{nmssmdm,Belanger:2005kh} in which $\lam$ and $\tanb\equiv v_u/v_d$ are positive while
$\kap$, $A_\lam$ and $A_\kap$ can have either sign.

\begin{figure}[tb]
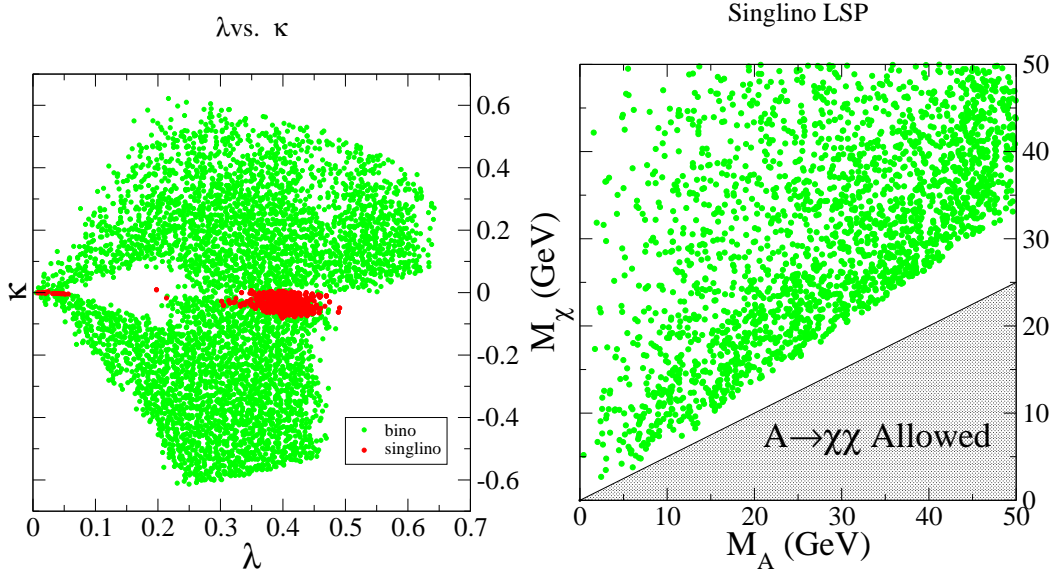

    \includegraphics[scale=0.3]{l_k.eps}
    \includegraphics[scale=0.3]{ma_mchi_singlino.eps}
    \caption{On the left, we show regions of $\lam$--$\kap$ parameter
space for which the $\cnone$ is singlino-like 
(defined by $\eps_s^2>0.5$) and bino-like 
(defined by $\eps_s^2\leq 0.5$). On the right, we plot $m_A$
    vs. $\mcnone$ for singlino-like 
    neutralinos with $\eps_s^2>0.9$. Each point shown is consistent with all LEP constraints.}
\label{l_k_chiovera}
\end{figure}

\section{Light Neutralinos in the NMSSM}\label{seclightchi}

\noi In the basis $\wtil\chi^0 = (-i\wtil\lam_1 , -i\wtil\lam_2, \psi_u^0, \psi_d^0,
\psi_s)$, the tree-level neutralino mass matrix takes the form
\beq
{\cal M}_{\wtil\chi^0} =
\left( \ba{ccccc}
M_1 & 0 & \frac{g_1 v_u}{\sqrt{2}} & -\frac{g_1 v_d}{\sqrt{2}} & 0 \\
0 & M_2 & -\frac{g_2 v_u}{\sqrt{2}} & \frac{g_2 v_d}{\sqrt{2}} & 0 \\
\frac{g_1 v_u}{\sqrt{2}} & -\frac{g_2 v_u}{\sqrt{2}} & 0 & -\mu & -\lam v_d \\
 -\frac{g_1 v_d}{\sqrt{2}} &  \frac{g_2 v_d}{\sqrt{2}} & -\mu & 0 & -\lam v_u \\
0 & 0 & -\lam v_d & -\lam v_u & 2 \kap x
\ea \right) . \eeq
In the above, the upper $4\times 4$ matrix corresponds to ${\cal
  M}_{\wtil\chi^0}^{\rm MSSM}$.  From the lower $3\times 3$ matrix, we
find that if $\lam v_{u,d}=(\mu/x)v_{u,d}$ are small compared to
$|\mu|$ and/or $2|\kappa x|$ then the singlino decouples from the MSSM
and has mass 
\beq
m_{\rm{singlino}} \simeq\sqrt{\lam^2v^2+4\kap^2 x^2}=\sqrt{\mu^2v^2/x^2+4\kap^2x^2}\,,
\label{msinglino}
\eeq
as found from $[M_{\wtil\chi^0}]^2_{55}$. Thus, if $2|\kap x|$ and
$\lam v$ are both $<M_1,M_2,|\mu|$, then the lightest neutralino will
tend to be singlino-like \cite{singlinopheno}.  Since $|x|$ is typically
substantial (given that $\lam<1$ and $|\mu|=\lam |x|$ must be substantial
to satisfy chargino mass limits), a singlino-like $\cnone$ (formally
defined by $\eps_s^2<0.5$) emerges mainly for small $\kap$.  In fact,
for very small $\lam$, $|x|$ must be quite large and thus the singlino
will be the LSP only if $|\kap|$ is also very small; otherwise
$2|\kap x|$ would exceed one or more of the typically moderate
$M_1,M_2,|\mu|$ values considered here and the singlino would not be
the LSP.  For larger $\lam$ (e.g. $\gsim 0.3$), $|x|$ need not be
extremely large and the singlino LSP condition
$2|\kap x|<M_1,M_2,|\mu|$ can hold for slightly larger $|\kap|$.  These
behaviors can be seen in Fig.~\ref{l_k_chiovera} obtained by scanning
using NMHDECAY 1.1~\cite{Ellwanger:2004xm}. NMHDECAY tests for theoretical
consistency of the model and for consistency with LEP constraints on
the Higgs sector, neutralinos and the chargino). It also includes
radiative corrections to the tree-level mass matrix that are often
quite important for small $|\kap|$. As expected, the neutralino is
singlino-like ($\epsilon_s^2>0.5$) when $|\kap|$ is small.  Consistent
solutions are found primarily in two regions of parameter space ---
one at small $\lambda$ with very small $|\kap|$, 
and another at large $\lambda$
with slightly larger $|\kap|$ allowed, see
Fig.~\ref{l_k_chiovera}.
Since $\tan\beta$ also induces singlino mixing,
the singlino points at large $\lambda$ also have small $\tan \beta
\lsim 4$ while the points at small $\lambda$ can have any value of
$\tan \beta$.

For $|\kap|$ not close to zero, bino-like $\cnone$'s can easily emerge
for small values of $M_1$.  In this case, the bino does not have a
large degree of mixing with the other neutralinos and the LSP mass is
nearly fixed to $M_1$.

We will find that a light
neutralino which is mostly bino or a combination of bino and singlino,
with a small admixture of higgsino, can generate the observed dark
matter density and evade all relevant collider constraints.

\section{Light CP-Odd Higgs Bosons in the NMSSM}\label{seclighta}

After removing the CP-odd degree of freedom that is absorbed in
giving the $Z$ its mass, the remaining CP-odd states have the
squared-mass matrix
\beq
{\cal M}_A^2=\left(\begin{array}{cc} {2\lam x \over \sin
      2\beta}(A_\lam+\kap x) & \lam v(A_\lam-2\kap x) \cr
\lam v(A_\lam-2\kap x) & (2\lam\kap+{\lam A_\lam\over 2 x})v^2\sin
2\beta-3x\kap A_\kap \cr\end{array}\right)
\eeq
where $v^2=v_u^2+v_d^2$. For physically acceptable solutions, the
lightest state must have $m_{A_1}^2>0$.  In addition, the lightest
CP-even Higgs boson and the charged Higgs boson must have positive
mass-squared. To avoid spontaneous CP-violation several other
conditions must be satisfied~\cite{nmssmdm}. In our conventions these
are
as follows.
\bit
\item For $\kap>0$, we must have one of three situations:
\ben
\item $\sign(\mu)=\sign(A_\lam)=-\sign(A_\kap)$;
\item $\sign(\mu)=-\sign(A_\lam)=-\sign(A_\kap)$ with $|A_\kap|>3\lam
v_uv_d |A_\lam|/(-|xA_\lam|+\kap x^2)$, where the denominator has to
be positive;
\item $\sign(\mu)=\sign(A_\lam)=\sign(A_\kap)$ with $|A_\kap|<3\lam
v_uv_d |A_\lam|/(|xA_\lam|+\kap x^2)$.
\een
\item For $\kap<0$, a CP-conserving minimum requires
\ben
\item
$\sign(\mu)=\sign(A_\lam)=\sign(A_\kap)$ with $|A_\kap|>3\lam
v_uv_d |A_\lam|/(|xA_\lam|-\kap x^2)$. 
\een
\eit

To find a model which has a light CP-odd Higgs boson, we can require
that one of the $U(1)_R$ or $U(1)_{PQ}$ symmetries approximately
holds. The $U(1)_R$ symmetry appears in the limit that the trilinear
terms $A_\kappa$ and $A_\lambda$ vanish.  This is well motivated from
models of gaugino-mediated SUSY breaking~\cite{gauginomediated} in
which trilinear terms are generated radiatively and, therefore, are
suppressed relative to the gaugino masses by a loop factor of $4\pi$.
One would expect $A_\lambda$ to be smaller than the gaugino masses by
a factor of $4 \pi$ and $A_\kappa$ to be smaller by a factor of $16
\pi^2$ because $S$ is not charged under gauge symmetries, and only
receives a trilinear term at two-loops.  The small trilinear terms are
also radiatively protected and remain small when evolved via RGEs from
the SUSY breaking scale to the weak scale \cite{nmssmaxion}.  In this
limit, the lightest CP-odd Higgs is a pseudo-goldstone boson of the
broken $U(1)_R$ symmetry and has a mass of $m_{A_1}^2 \simeq -3 \kappa
A_\kappa x$ in the large $\tan \beta$ or large $|x|$ limits.
Alternatively, we can make the substitution, $x =\mu/\lambda$, and
write this as 
$m_{A_1}^2 \simeq -3 \frac{\kappa}{\lambda} A_\kappa \mu$. 

More generally, in the limit of small $A_\lam$ and $A_\kap$ one finds
\beq
\tan\theta_A\sim {x\over v\sin 2\beta}\,,\quad
\cos^2\theta_A\sim{v^2\sin^22\beta\over v^2\sin^22\beta +x^2}\,,
\eeq
and 
\beq
m_{A_1}^2\sim {{9\over 2}\lam A_\lam v^2 x \sin2\beta-3\kap A_\kap
  x^3\over x^2+v^2\sin^22\beta}\,.
\eeq 
Since $|x|>v$ is preferred, $|\cos\theta_A|$ is typically small 
at small to moderate $\tanb$,
with $\cos^2\theta_A\to 0$ at large $\tanb$. 
If we only take $A_\kap\to 0$, one finds the results
\beq
\tan\theta_A\sim {x\over v\sin2\beta}\left[{1+A_\lam/(\kap x) \over
  1-A_\lam/(2\kap x)}\right]\,,
\eeq
and
\beq
\label{ma1largeak}
m_{A_1}^2\sim {{9\over 2}\lam A_\lam v^2 x \sin2\beta
  \over x^2+v^2\sin^22\beta+A_\lam x/\kap}\,,
\eeq 
valid whenever the numerator of the preceding
equation is much smaller than the square
of the denominator, as for example if $\tanb\to \infty$ or $x$
is large. Again, $\cos\theta_A$ will be quite small typically
and the $A_1$ relatively singlet like.
In practice, this limit is very frequently applicable.

The $U(1)_{PQ}$ symmetry appears in the limit that $\kappa$
vanishes (and therefore the soft SUSY breaking term $\kappa A_\kappa$
also vanishes) and also results in a light $A_1$.  To leading order,
one finds
\beq
\tan\theta_A\sim -{2x\over v\sin 2\beta}\,,\quad \cos^2 \theta_A\sim
{v^2\sin^22\beta\over v^2\sin^22\beta+4x^2}
\eeq
and 
\beq m_{A_1}^2\sim {6\kap x^2(3\lam v^2\sin2\beta-2A_\kap
  x)\over 4x^2+v^2\sin^22\beta}\,. 
\eeq 
For $|x|>v$, $|\cos\theta_A|$ is small
for moderate $\tanb$ and approaches $0$
at large $\tanb$.  

It is useful to note that if $\lam$ is small (implying large $|x|$),
then singlet mixing in both $M_{\wtil \chi_0}$ and ${\cal M}_A^2$
is small. If $2|\kap x|<M_1,M_2,|\mu|$
and $2\mu(A_\lam+\kap x)/\sin2\beta>-3\kap A_\kap x$
then the $\cnone$ and $A_1$ will
both be singlet in nature. In particular, 
for small $|\kap|$ both the $A_1$ and the
$\cnone$ can easily be singlet-like.  
At large $\lambda$, the $A_1$ can have a more mixed
nature ($\cos \theta_A$ tends to be larger) but as we have seen
a moderately-singlino $\cnone$
 is still allowed despite the somewhat
larger mixing in the neutralino mass matrix.

In either of the $A_\lam,A_\kap\to 0 $ or $\kap\to 0$ cases, a light
$A_1$ is technically natural since it is protected by an approximate
symmetry.  From an effective field theory perspective, the small terms
that break the symmetry will not receive large radiative corrections.
It is technically natural for the $A_1$, $H_1$ or $\cnone$ to be very
light as a result of $U(1)_R$ and/or $U(1)_{PQ}$ symmetries.

The fermion Yukawas break the $U(1)_R$ symmetry, leading to
contributions arising at one-loop for the $H_u$ and $H_d$ components of
the Higgs sector, and at two-loops for the $S$ component.  However the
radiative corrections to the singlet component are proportional to
either $\lambda$ or $\kappa$, and thus are suppressed for small values
of $\lambda,\kappa$. These symmetries therefore result in the hierarchy
$m_h \gg A_\lambda \gg A_\kappa$.

It will be helpful in understanding dark matter relic
density issues to examine whether or not a light
(singlet) $A_1$ can decay to a pair of nearly pure $\cnone$
singlinos. From the right hand
plot of Fig.~\ref{l_k_chiovera}, we observe that it 
is impossible to
obtain $m_{A_1} > 2 \mcnone$ when the LSP is nearly purely singlino
($\eps_s^2>0.9$).  Using the fact that a highly singlino
$\cnone$ is achieved by taking $\lam$ to be very small
(so as to remove mixing in $M_{\wtil \chi_0}$),
implying large $|x|$, we are able to analytically
understand this in
two cases: (i) $|\kap x|$ moderate in size and (ii)
$|\kap x|$ small.  For moderate $|\kap x|$, 
the inability
to satisfy the mass requirements for the decay $A_1 \rightarrow
\cnone\cnone$ stems from the inability to simultaneously
satisfy $m_{S_3}^2>0$ and $m_{A_1}^2/4 \mcnone^2>1$. Here, $S_3$ is
the third CP-even (largely singlet) Higgs mass eigenstate as defined in~\cite{nmssmus}
and is the lightest CP-even Higgs state in the limit of interest.
For small $\lam$ and finite $|\kap x|$, we have 
\beq
\mcnone\sim 2\kap x\,,\quad m_{A_1}^2\sim -3 \kap A_\kap x\,,
\eeq
the latter requiring $\kap A_\kap <0$.
Further, if one expands the CP-even mass matrix in the large $|x|$
limit, holding $\mu$ and $\kap x$ fixed, one finds~\cite{nmssmus} (after
correcting for differences in sign conventions)
\beq
m_{S_3}^2=4\kap^2x^2+\kap A_\kap x+{\mu^2 v^2\over \kap^2
  x^2}\left[{\mu\over x}-{1\over 2}\left(-2\kap +{A_\lam\over x}\right)
\sin 2\beta\right]^2+{\mu^2v^2\left(-2\kap +{A_\lam\over
      x}\right)^2\cos^22\beta\over 4\kap^2x^2-{2\mu x A_\Sigma\over \sin2\beta}}
\eeq
where $A_\Sigma\equiv A_\lam-\kap x$. For fixed $|\kap x|$ with $|x|$ very
large, the last two terms approach zero and we have
$m_{S_3}^2\sim 4\kap^2x^2+\kap A_\kap x$ which is positive only if
$-\kap A_\kappa < 4 \kappa^2 x$. However, in the same limit, the mass
condition for the $A_1 \rightarrow \cnone\cnone$,
written as $m_{A_1}^2>4\mcnone^2$, becomes $-\kap A_\kappa > (16/3) \kappa^2 x$.  These
two conditions cannot be simultaneously satisfied, and thus the decay
$A_1 \rightarrow \cnone \cnone$ is not allowed for a pure singlino
in the large $|x|$, fixed $\kap x$ limit.
Some admixture of bino, and therefore moderate $\lambda$ and $|x|$ must
be required for this decay to be open.
                                                                                
The $|\kap|\to 0$ case (the Peccei-Quinn symmetry limit) at small $\lam$
is defined by
\begin{equation}
    |\kap| \ll \mathcal{O}\left( \lambda, \frac{|A_\lambda|}{v}, \frac{|A_\kappa|}{v},
    \frac{|\mu|}{|x|}, \frac{v}{|x|} \right), \qquad v \ll |x|.
\end{equation}
In this limit, Eq.~(\ref{msinglino}) implies $\mcnone^2 \simeq
\lambda^2 v^2=\mu^2 v^2/x^2$. Meanwhile, for $|\kap x|\to 0$ and $|x|$
large ($\lam $ small), it is easily seen that
$m_{A_1}^2\propto 1/|x|^3$.  Thus, once again,
the $A_1\to\cnone\cnone$ decay is disallowed.

The fact that $m_{A_1}<2\mcnone$ in these singlino limits implies
that a singlino $\cnone$ is disfavored cosmologically.
This is because $m_{A_1} \simeq 2 \mcnone$ is required to
enhance the annihilation cross section to the level needed to obtain the correct
relic density.  Therefore some bino mixing is required to get an
appropriate relic density.  This, in turn, requires $M_1$ to be small as
well.


Finally, we should note that 
the constraints on a light $A_1$ are rather weak. This
is because direct searches at LEP~\cite{lepabounds} require a light $A_1$
to be radiated off of a quark or a tau lepton and, due to the small fermion
Yukawa couplings, bounds are only obtainable when the $A_1$ coupling to fermions
is enhanced by $\tan \beta$.  However, in the $U(1)_{PQ}$ and $U(1)_R$
symmetry limits discussed earlier, $\cos\theta_A$ (the non-singlet
part of $A_1$) is proportional to $\sin 2\beta$  so that
the product $\tan\beta\cos\theta_A$ remains modest in size.
 If a light $A_1$ exists and is near in mass to the $\eta$ (547 MeV),
    it may be discovered via invisible decays of the $\eta$ at low
    energy lepton colliders.~\cite{bottomonium}.  This mass range is
    extremely interesting if a light $A_1$ exchange is the explanation
    for the recent galactic 511 keV line from the INTEGRAL/SPI
    experiment~\cite{511annihilate}.

We now give some additional remarks concerning the singlet and
non-singlet
$A_1$ possibilities.

\subsection{Models with a Singlet-Like $A_1$}

As we have seen, a singlet-like $A_1$ ($\cos \theta_A \ll 1$) is
extremely easy to obtain by 
making some combination of $|\kappa|$, $|A_\kappa|$ and $|A_\lambda|$
small. 
Indeed, the mass of the  $A_1$ can be driven to zero at tree level.
Radiative corrections increase this mass, however.  The dominant source
of these radiative corrections to the singlet mass is from Standard Model
couplings since the number of degrees of freedom is much larger in the
MSSM than in the singlet supermultiplet.  These radiative corrections
are therefore proportional to $\lambda$, since the $\lambda$
superpotential term is the {\it only} coupling connecting the singlet
with the rest of the MSSM.  Therefore, if the light singlet mass is to be
radiatively stable, $\lambda$ must be small.  $\lambda$ being small also has
the effect of reducing the mixing with the singlet component in both the
CP-even mass matrix and the neutralino mass matrix.  {\it All} terms
which mix the singlino with the MSSM neutralinos and the singlet $S$
with $H_u$ and $H_d$ are proportional to $\lambda$.  We find $\lambda
\lsim 0.1$ to be natural, with larger values of $\lambda$ requiring an
increasing amount of cancellation between the various radiative
contributions to its mass.  $\lambda$ being this small necessarily implies
that the singlet vacuum expectation value, $|x|$, is large since
$\mu=\lambda x$.  Chargino searches generally imply
 $|\mu| \gsim 100$ GeV, leading to
 $|x| \gsim 1$ TeV for $\lam\lsim 0.1$.  Furthermore, with all four of
$\lambda$, $\kappa$, $A_\kappa$ and $A_\lambda$ small
in magnitude, the entire
supermultiplet is light and $A_1$, $H_1$ (largely the singlet-like
$S_3$)
and $\cnone$ tend to be nearly degenerate.

\subsection{Models with an MSSM-Like $A_1$}

An MSSM-like (non-singlet) $A_1$ ($\cos \theta_A \simeq 1$) can also be obtained, but is
subject to more stringent constraints. If $\cos\theta_A\simeq 1$, 
couplings of the $A_1$ to down-type fermions go like $\cos \theta_A \tan
\beta$, therefore phenomenological constraints become significant at large
$\tan \beta$. If such an $A_1$ is very light, it will be further constrained by rare decays such as $K \rightarrow \pi \nu
\bar{\nu}$ and $\Upsilon \rightarrow \gamma X$ as discussed in Sec.~\ref{upsconstraints}.

\section{Constraints}

In this section, we will consider a series of constraints which may be
relevant to light neutralinos and/or a light CP-odd Higgs bosons in the NMSSM.
Except for the LEP and $\Upsilon$ decay limits, most of
the constraints discussed below are easily avoided
by appropriate choices of SUSY parameters to which our
dark matter calculations are not sensitive.

\subsection{LEP Limits}

If the lightest neutralino is lighter than $m_Z/2$, $Z$ decays to
neutralino pairs may violate the bounds obtained at LEP for the $Z$'s
invisible decay width. In particular, we require $\Gamma_{Z \rightarrow
\cnone \cnone} < 4.2$ MeV, which corresponds to one standard deviation
from the measured neutrino contribution. Since binos, winos and
singlinos do not couple to the $Z$, this constraint can only limit the
higgsino components of the lightest neutralino. In the mass range we are
most interested in here ($m_{\cnone} \lsim 20$ GeV), this constraint is
satisfied for all models with $| \epsilon_u^2 - \epsilon_d^2 | \lsim 6
\%$.  

Direct chargino searches also limit the wino component of the lightest
neutralino.  This is due to the fact that
if the lightest neutralino has a significant
wino component, then it will have a mass that is a significant
fraction of the chargino mass (with near degeneracy if
the $\cnone$ is mainly wino).

In combination, these constraints imply that a very light $\cnone$
must be dominantly a linear combination of bino and singlino.

\subsection{The Magnetic Moment of the Muon}

The one-loop contribution to the magnetic moment of the muon from a light neutralino comes from a triangle diagram with a smuon along two sides and the neutralino around the third. This contribution is given by \cite{gminus2neutralino}:
\begin{equation}
\delta a_{\mu}^{\cnone} = \frac{m_{\mu}}{16 \pi^2} \sum_{m=1,2} \bigg[
\frac{- m_{\mu}}{12 m_{\tilde{\mu}_m}^2} (|n_m^L|^2 + |n_m^R|^2)
F_1^N(x_m) + \frac{m_{\cnone}}{3 m_{\tilde{\mu}_m}^2} {\rm Re}[n_m^L n_m^R] \,  F_2^N(x_m) \bigg],
\end{equation}
where $n_m^R = \sqrt{2} g_1 \epsilon_B X_{m2} + y_{\mu} \epsilon_u X_{m1}$, $n_m^L = (g_2 \epsilon_W X_{m2} + g_1 \epsilon_B) X^*_{m1}/\sqrt{2}-y_{\mu} \epsilon_u X^*_{m2}$, $y_{\mu}=g_2 m_{\mu}/(\sqrt{2}m_W \cos \beta)$ and $X_{m,n}$ are elements of the unitary matrix which diagonalizes the smuon mass matrix. The functions, $F^N$, are defined by:
\begin{eqnarray}
F^N_1 (x_m)&=& \frac{2}{(1-x_m)^4} \bigg(1-6x_m+3x_m^2+2x_m^3-6x_m^2 \ln x_m \bigg), \\
F^N_2 (x_m)&=& \frac{3}{(1-x_m)^3} \bigg(1-x_m^2+2x_m \ln x_m \bigg),
\end{eqnarray}
where $x_m = m^2_{\cnone}/m^2_{\tilde{\mu}_m}$. For a bino-like neutralino, this reduces to:
\begin{equation}
\delta a_{\mu}^{\cnone} \simeq \frac{g_1^2}{48 \pi^2} \frac{m_{\mu} m_{\cnone}}{(m^2_{\tilde{\mu}_2}  - m^2_{\tilde{\mu}_1})} \Delta_{\tilde{\mu}} \bigg[ \frac{F_2^N(x_1)}{m^2_{\tilde{\mu}_1}} - \frac{F_2^N(x_2)}{m^2_{\tilde{\mu}_2}}  \bigg],
\end{equation}
where $\Delta_{\tilde{\mu}}$ is the real part of the off diagonal elements of the smuon mass matrix, $\Delta_{\tilde{\mu}} = \Re[m_{\mu}(A_{\tilde{\mu}}-\mu^* \tan \beta)]$. This further simplifies to:
\begin{equation}
\delta a_{\mu}^{\cnone} \sim  2.3 \times 10^{-11} \bigg(\frac{m_{\cnone}}{10 \, \rm{GeV}}\bigg) \bigg(\frac{200 \, \rm{GeV}}{m_{\tilde{\mu}}}\bigg)^4 \bigg(\frac{\mu \tan \beta - A_{\tilde{\mu}}}{1000 \, \rm{GeV}} \bigg).
\label{g2chi}
\end{equation}

In addition to this contribution from a light neutralino, a light CP-odd Higgs can contribute non-negligibly to $\delta a_{\mu}$ through both one-loop and two-loop processes \cite{gminus2higgs}. The one-loop contribution (corresponding to a triangle diagram with a muon along two sides and the CP-odd Higgs along the third side) is given by: 
\begin{equation}
\delta a_{\mu}^{A\, 1\,\rm{loop}} = \frac{g_2^2 \, m^2_{\mu} \, \cos^2 \theta_A  \tan^2 \beta \, L_A}{32 \, m^2_W \, \pi^2}, 
\end{equation}
where the function, $L_A$, is given by:
\begin{equation}
L_A = \frac{m_{\mu}^2}{m_A^2} \int^1_0 \frac{-x^3 \, dx}{x^2 \,(m_{\mu}^2/m_A^2) + (1-x)}.
\end{equation}
Numerically, this function yields $L_A=-0.032$, $-0.00082$ and $-0.000014$ for 1, 10 and 100 GeV CP-odd Higgs bosons, respectively. Thus we arrive at:
\begin{equation}
\delta a_{\mu}^{A\, 1\,\rm{loop}} \approx L_A \cos^2 \theta_A  \tan^2 \beta \times 2.7 \times 10^{-9},
\end{equation}
which is a negative contribution due to the sign of $L_A$.

The contribution from two-loop diagrams involving a heavy fermion loop is given by:
\begin{equation}
\delta a_{\mu}^{A \, 2\,\rm{loop}} = \frac{g_2^2 \, m^2_{\mu} \, \alpha \, c_f \, Q_f^2 \, \chi_f^2 \, L_f}{32 \, m^2_W \, \pi^2}, 
\end{equation}
where $c_f$ is the color factor of the fermion in the loop (3 for quarks, 1 for
leptons), $Q_f$ is the electric charge of the fermion, $\alpha \approx 1/137$,
$\chi_f= \cos \theta_A \tan \beta$ for up-type fermions and $\cos \theta_A \cot
\beta$ for down-type fermions and $L_f$ is given by:
\begin{equation}
L_f = \frac{m_f^2}{2 m_A^2} \int^1_0 \frac{dx}{x \,(1-x) -(m_f^2/m_A^2)} \, \ln\bigg(\frac{x\, (1-x)}{m_f^2/m_A^2}\bigg).
\end{equation}
For a top quark loop, this function yields the values $L_t =$ 6.2, 3.9 and 1.7 for 1, 10 and 100 GeV CP-odd Higgs bosons, respectively. For a bottom quark loop, these values are $L_b =$ 2.5, 0.59 and 0.038. Numerically, these contributions are:
\begin{equation}
\delta a_{\mu}^{A \, 2\,\rm{loop}} \approx L_t \,  \cos^2 \theta_A \, \cot^2 \beta \times 2.6 \times 10^{-11} + L_b \,  \cos^2 \theta_A \,\tan^2 \beta \times 6.6 \times 10^{-12}.
\end{equation}
Combining the results of the one and two-loop contributions from the light
$A_1$, we arrive at the following:
\begin{eqnarray}
\delta a_{\mu}^{A \, 1+2\,\rm{loop}} \approx -7 \times 10^{-11} \times \cos^2 \theta_A \tan^2 \beta \,\,\,\,\,\,\,\, \mbox{for} \,\, m_{A_1} &=& 1 \, \rm{GeV},\nonumber \\ 
\delta a_{\mu}^{A \, 1+2\,\rm{loop}} \approx 1.7 \times 10^{-12} \times \cos^2 \theta_A \tan^2 \beta \,\,\,\,\,\,\,\,  \mbox{for}\, \, m_{A_1} &=& 10\, \rm{GeV}.
\end{eqnarray}

It is somewhat difficult to know how best to interpret the current status of the measurement of the muon's magnetic moment. Using $e^+ e^-$ data, the measured value exceeds the theoretical prediction by $\delta a_{\mu} (e^+e^-)=[23.9 \pm 7.2_{\rm{had-lo}}  \pm 3.5_{\rm{lbl}}  \pm 6_{\rm{exp}}] \times 10^{-10}$, where the error bars correspond to theoretical uncertainties in the leading order hadronic and the hadronic light-by-light contributions as well as from experimental contributions. Combined, this result is 2.4$\sigma$ above the Standard Model prediction. Experiments using $\tau^+ \tau^-$ data, on the other hand, find $\delta a_{\mu} (\tau^+ \tau^-)=[7.6 \pm 5.8_{\rm{had-lo}}  \pm 3.5_{\rm{lbl}}  \pm 6_{\rm{exp}}] \times 10^{-10}$, which is only 0.9$\sigma$ above the Standard Model prediction \cite{gminus2data}.

Comparing the expression shown in Eq.~(\ref{g2chi}) for a light neutralino to
these experimental results illustrates that only in extreme models, with a
combination of small $m_{\tilde{\mu}}$, large $\tan \beta$ and large $\mu$ is
there any danger of exceeding these bounds with a light neutralino. For fairly
moderate choices of parameters, {\it i.e.} $m_{\tilde{\mu}} \sim 200$ GeV, $\tan
\beta \sim 20$ and $\mu \sim 500$ GeV, the experimental values can be
matched. Since $m_{\tilde{\mu}}$ is not currently known and
the dark matter scenarios we consider are not sensitive to
$m_{\tilde{\mu}}$, we will not 
consider this constraint in our dark matter calculations.

The contribution from a light CP-odd Higgs also should not violate
the $\delta a_\mu$ constraint.
If one considered the $A_1$ contribution alone, one might conclude that $\cos
\theta_A \tan \beta $ is strongly limited in the case of
small $m_{A_1}$. However, contributions to $\delta a_\mu$ from other
sources such as the charged Higgs, charginos, and sfermions can easily overwhelm
or cancel any contribution from a light $A_1$.  Furthermore,
LEP and other indirect limits such as $\Upsilon$ decays (discussed in
Sec.\ref{upsconstraints}) constrain $\cos \theta_A
\tan \beta$ to be small, so it is generally not possible to see a large
enhancement in $\delta a_\mu$.
Finally we note that if the $A_1$ and $\cnone$
are both light, as considered here, their contributions
to $\delta a_\mu$ are of opposite sign, and can cancel.

Thus, we do not explicitly include the $\delta a_\mu$ constraints
in our computations. 
Their inclusion would only become appropriate if a specific 
model for soft-SUSY-breaking is being considered.


\subsection{Rare Kaon Decays}

The $K^+\rightarrow \pi^+ \nu \bar{\nu}$ branching ratio was recently
measured by the E787 and E949 experiments to be $BR(K^+\rightarrow \pi^+ \nu \bar{\nu})=(1.47^{+1.30}_{-0.89})\times
10^{-10}$, which is nearly twice the value predicted in the Standard Model,
$(0.67^{+0.28}_{-0.27})\times 10^{-10}$~\cite{kpinunu}. A CP-even Higgs boson lighter than a few hundred MeV can contribute to
this branching ratio via a triangle diagram involving $W^\pm$ bosons on
two sides, and an up or charm quark on the third.  This contribution is
suppressed by $\xi_u$ and $\xi_d$. CP-odd Higgs bosons, on the other hand, cannot contribute to this process at the one loop level since
the vertex involving $W$'s and the $A_1$ is $W^\mu W_\mu A_1 A_1$ and, therefore,
the leading contribution to $K^+\rightarrow \pi^+ +$ invisible has {\it
four} $\cnone$'s in the final state.  This requires a $\cnone$ lighter
than 88.5 MeV, which is lighter than the range we consider in this study.  

Other rare kaon decays such as $K^0\rightarrow e^+ e^-$ and $K^+\rightarrow
\pi^+ e^+ e^-$ are similarly unconstraining for a light $A_1$, but
potentially important for a light $H_1$ for the same reasons.

A recent study~\cite{Hiller:2004ii} analyzed this in detail and concluded that
extremely light $m_{A_1} < 2 m_\mu$ can be ruled out. However, this can be evaded
if $|\kappa|$ is small enough.

\subsection{Rare $B$-Meson Decays}

The transitions $b\rightarrow s \gamma$ and $b \rightarrow \mu^+ \mu^-$
are usually considered sensitive probes of supersymmetry, however both
are flavor changing, while a light $A_1$ and $\chi^0_1$ are not flavor
changing by themselves.  These and other flavor changing processes
involving a light $\chi^0_1$ propagator can always be suppressed by
making the appropriate squark or slepton mass heavy since the relevant
diagrams must involve a $f\tilde{f}\chi^0_1$ vertex.  Processes
involving a light $A_1$ may be suppressed at one-loop by assuming the
Minimal Flavor Violation mechanism~\cite{Hiller:2004ii,Bobeth:2001sq}.
A recent study of the $B$ meson decays $b\rightarrow s \gamma$, $b
\rightarrow s A_1$, and $b \rightarrow s l^+ l^-$  in the NMSSM
concluded that $A_1$ masses down to $2 m_e$ cannot be excluded from
these constraints~\cite{Hiller:2004ii}.

Another rare $B$-decay is $B^+ \rightarrow K^+ \nu \bar{\nu}$.  This
process also necessarily involves a quark flavor-changing $W^\pm$
vertex.  A diagram in which the light $A_1$ couples to the $W^\pm$ must
involve {\it two} $A_1$'s and {\it two} $W^\pm$'s unless CP is violated,
severely limiting the set of processes to which it can contribute.
Diagrams where the light $A_1$ couples to the fermion also must have a
$W^\pm$ to change the quark flavor and also receive a factor of $\cos
\theta_A$ at each $f \bar{f} A_1$ vertex, strongly suppressing the $A_1$
contribution for the scenarios we focus on, all
of which have small $\cos\theta_A$.

\subsection{Upsilon and $J/\Psi$ Decays}
\label{upsconstraints}

The vector resonances $J/\Psi$ and $\Upsilon$ may decay radiatively into
an $A_1$ and a photon if $A_1$ is sufficiently light.  There are two experimental limits on this process: firstly when the $A_1$ decays invisibly or is long-lived enough to leave the
detector volume~\cite{ups1stogammainvis}, and secondly when the $A_1$ decays to Standard Model particles~\cite{Besson:1985xw}.
This width, relative to the width to muons at leading order is given by~\cite{upsaxion}: 
\begin{equation}
    \label{vaxiondecay}
    \frac{\Gamma(V \rightarrow \gamma A_1)}
    {\Gamma(V \rightarrow \mu \mu)} =
    \frac{G_F m_b^2}{\sqrt{2}\alpha
    \pi}\left(1-\frac{M_{A_1}^2}{M_V^2}\right)  X^2,
\end{equation}
where $V$ is either $J/\Psi$ or $\Upsilon$ and
$X=\cos\theta_A\tan\beta$ for 
$\Upsilon$ and $X=\cos\theta_A\cot \beta$ for $J/\Psi$.  The $A_1$ is
often referred to as the axion in this literature.
Eq.~(\ref{vaxiondecay}) is also applicable for a light CP-even $H_1$,
with $X=\xi_d/\cos \beta$ for $V=\Upsilon$ and $X=\xi_u/\sin\beta$
for $V=J/\Psi$.

It is usually claimed that a light MSSM $A$ is ruled out if it is light enough so that both the
$\Upsilon$ and $J/\Psi$ can decay to it ($m_A \lsim 3.1$ GeV), due to the observation that
$\Gamma(J/\Psi\rightarrow \gamma A) \times \Gamma(\Upsilon \rightarrow
\gamma A)$ is independent of $\tan \beta$.  However, within the NMSSM, this product is proportional to $\cos^4 \theta_A$,
which may be small.

The best existing measurement of $\Upsilon\to invisible+\gam$ is from
CLEO~\cite{ups1stogammainvis} in 1995.  Significantly more data
has been collected on the $\Upsilon(1S)$ resonance that could be used to
improve this measurement, however.  Modern $B$-factories such as BaBar and Belle
can also produce the $\Upsilon(1S)$ in Initial State Radiation to
improve this measurement.

\begin{figure}[tb]
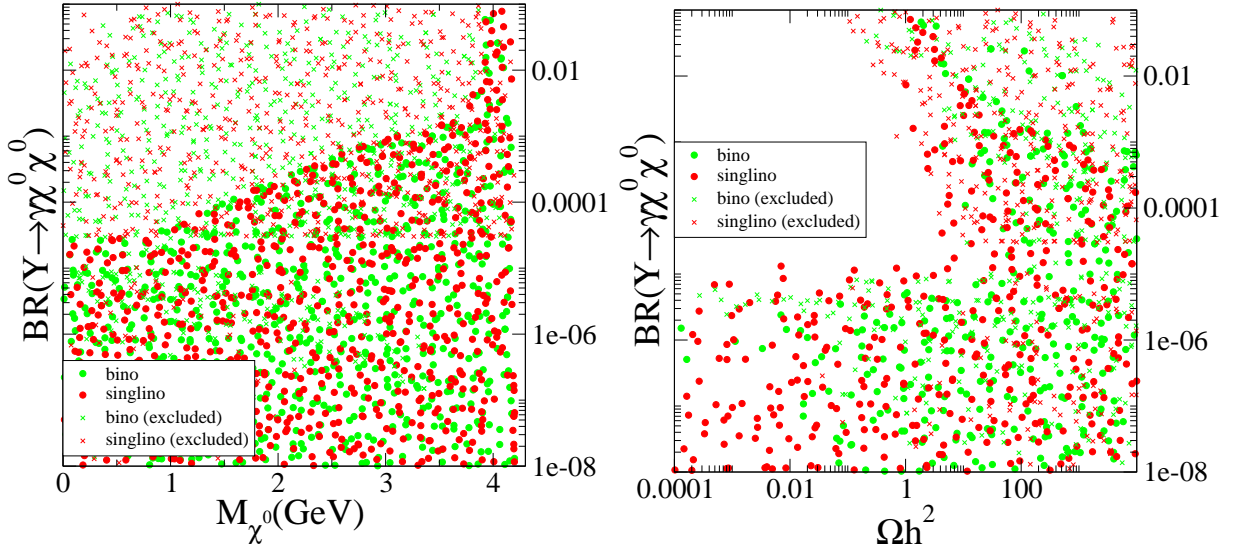

    \includegraphics[width=8cm]{mn_brups.eps}
    \includegraphics[width=8cm]{omegah2_brups.eps}
\caption{The branching ratio for $\Upsilon(1S)\to \gam\cnone\cnone$  
via 3-body decay (i.e. either $m_{A_1}<2\mcnone$ or
$m_{A_1}>m_\Upsilon$) 
is plotted vs.
the LSP mass (left) and relic density $\Omega h^2$ (right). All points shown are
consistent with all LEP constraints.  Points marked by an x are excluded by
one of: $\Upsilon \rightarrow \gamma \cnone \cnone$ (3-body decay)
(that which is plotted); $\Upsilon \rightarrow
\gamma A_1$ (2-body decay)
with  $A_1 \rightarrow \cnone \cnone$
(2-body decay);  or
$\Upsilon \rightarrow \gamma A_1$ (2-body decay) where the $A_1$ decays visibly.
}

\label{mn_brups}
\end{figure}

In Fig.~\ref{mn_brups}, we illustrate the correlations between
$\br(\Upsilon\to\gam\cnone\cnone)$ via 3-body decay (i.e. {\it not} 
$\Upsilon\to \gam A_1$ with $A_1\to invisible$ or $visible$),
$\mcnone$ and the relic 
density $\Omega h^2$ (the calculation of which is discussed
in the following section). 
Only higgs exchange is included in these relic density values.
Sub-leading $Z$ and sfermion exchanges would further decrease the relic
density of points with very large $\Omega h^2$.
The left-hand plot shows
that a significant fraction of the parameter
choices such that $\Upsilon\to\gam\cnone\cnone$ is allowed
are eliminated by the experimental constraint on this mode,
with additional ones being eliminated by the constraints
on the 2-body $\Upsilon\to\gam A_1$ decay mode. But, many
are not excluded, especially those with a bino-like $\cnone$.
Improvement in the experimental sensitivities to 
$\br(\Upsilon\to\gam\cnone\cnone)$ and $\br(\Upsilon\to\gam A_1)$ 
will further constrain the light
$\cnone$ scenarios considered here, 
or could yield a signal. The right-hand plot of Fig.~\ref{mn_brups}
shows that there are many parameter choices
that yield $\br(\Upsilon\to\gam\cnone\cnone)$ 
and $\br(\Upsilon\to\gam A_1)$ below the experimental
limits while simultaneously predicting
a relic density roughly consistent with observation.
We observe that this dual consistency can achieved for either a
bino-like or a singlino-like lightest neutralino.
 
The points on the left side of the right frame of Fig.~\ref{mn_brups} undergo $\Upsilon \rightarrow \gamma \cnone \cnone$ dominantly via a CP-even, mostly-singlet scalar, $H_1$, which mediates this
interaction.  When the $A_1$ becomes light and mostly singlet, it often
brings the CP-even scalar and singlino down in mass as well. For these points, the two body decay, $\Upsilon
\rightarrow \gamma A_1$, followed by the decay, $A_1 \rightarrow \cnone
\cnone$, is also just below the experimental limit.

\section{Annihilation Cross Section and Relic Abundance}

The calculation of the neutralino annihilation cross section and relic
abundance in the NMSSM is only slightly modified from the case of the
MSSM. First, the diagonalization of the $5\times5$ neutralino mass
matrix of the NMSSM yields different LSP compositions for given choices
of input parameters ($M_1$, $\mu$, etc.). Secondly, annihilations can
occur through the exchange of a Higgs boson with a significant singlet
component. On one hand, this weakens the respective couplings.  On the
other hand, much lighter Higgses can be considered, as collider
constraints are weakened.

For the range of masses we are considering, the only final states
available for the annihilations of light neutralinos are fermion pairs.
This process can occur through $s$-channel Higgs exchange (both CP-even
and CP-odd), $s$-channel $Z$ exchange or $t$ and $u$-channel sfermion
exchange.  With LEP constraints limiting sfermion masses to
$m_{\tilde{f}} \gsim 100$ GeV, neutralinos lighter than approximately 25
GeV cannot annihilate efficiently enough through sfermions to yield
the measured relic density. Similarly, $Z$ exchange cannot dominate the
annihilation cross section for light neutralinos. Therefore, we focus on
the process of Higgs exchange.

The squared amplitudes for the processes, $\cnone \cnone \rightarrow A \rightarrow f \bar{f}$ and  $\cnone \cnone \rightarrow H \rightarrow f \bar{f}$, averaged over the final state angle are given by \cite{amp}:
\begin{eqnarray}
\omega^A_{f\bar{f}} &=& \frac{C^2_{f\anti fA} \, C^2_{\cnone \cnone A}}{
  (s-m^2_A)^2 +  m^2_A \Gamma^2_A  } \, \frac{s^2}{16 \pi} \sqrt{1 -
  \frac{4 m^2_f}{s}}, 
\label{crossx1}\\
\omega^H_{f\bar{f}} &=& \frac{C^2_{f\anti fH} \, C^2_{\cnone \cnone H}}{
  (s-m^2_H)^2 +  m^2_H \Gamma^2_H} \, \frac{(s-4
  m^2_{\cnone})(s-4m^2_f)}{16 \pi} \sqrt{1 - \frac{4 m^2_f}{s}},
\label{crossx2}
\end{eqnarray}
where the labels $A$ and $H$ denote a CP-odd and CP-even Higgs,
respectively. Here, $C^2_{f\anti fA}$,  $C^2_{f\anti fH}$, $C^2_{\cnone \cnone A}$ and
$C^2_{\cnone \cnone H}$ are the fermion-fermion-Higgs couplings and the
neutralino-neutralino-Higgs couplings, and $m_{A, H}$ and $\Gamma_{A,H}$ are
the Higgs masses and widths. In the NMSSM case, we will
be considering only $A=A_1$ and $H=H_1$, the lightest of
the CP-odd and CP-even states, respectively. The relevant couplings
are then given by:
\begin{eqnarray}
    C_{\cnone\cnone A} &=& \cos \theta_A \left[
        (g_2 \epsilon_W - g_1 \epsilon_B)(\epsilon_d \cos \beta - \epsilon_u \sin\beta)
        +\sqrt{2}\lambda \epsilon_s (\epsilon_u \sin \beta + \epsilon_d \cos \beta)
        \right] \\ \nonumber
        &+& \sin \theta_A\sqrt{2}\left[\lambda\epsilon_u \epsilon_d
        - \kappa \epsilon_s^2\right] \\ \nonumber \\ 
    C_{\cnone\cnone H} &=& 
        (g_1\epsilon_B-g_2\epsilon_W)(\epsilon_d\xi_u - \epsilon_u\xi_d)
        + \sqrt{2}\lambda\epsilon_s(\epsilon_d\xi_d+\epsilon_u\xi_u) 
        + \sqrt{2}\xi_s(\lambda\epsilon_u\epsilon_d-\kappa\epsilon_s^2) \\ \nonumber \\
    C_{f\anti f A} &=& \left\{
\begin{array}{ll} \frac{m_f}{\sqrt{2}v}\cos\theta_A \tan \beta, &
  f=d,s,b,l \cr \frac{m_f}{\sqrt2v}\cos\theta_A\cot\beta, & f=u,c
  \cr\end{array}\right.
  \\ \nonumber \\
    C_{f\anti f H} &=& \left\{
\begin{array}{ll} \frac{m_f}{\sqrt{2}v}\frac{\xi_d}{\cos\beta}, &
f=d,s,b,l \cr \frac{m_f}{\sqrt2v}\frac{\xi_u}{\sin\beta}, & f=u,c\,.\cr
\end{array}\right.
\end{eqnarray}

We expect $\Gamma_{A} \approx$ eV-MeV if $A=A_1$ is mostly singlet and
$\Gamma_{A} \approx$ 1-10 MeV otherwise. Similarly, we expect $\Gamma_{H} \approx$ 10 eV-100 keV if $H_1$ is
mostly singlet and $\Gamma_{H} \approx$ keV-MeV if
$H=H_1$ is mostly non-singlet.  These widths are strongly affected by
the many kinematic 
thresholds due to hadronic resonances with masses less than 10 GeV.
Therefore, any computation of the relic density is inherently limited by
our ability to compute hadronic form factors and sum over hadronic
decays which may be on-shell and may enhance the annihilation.  We
require only that the relic density is $\mathcal{O}(0.1)$.  There is
sufficient parameter space to make the relic density precisely the value
measured by WMAP when all hadronic corrections are taken into account.
In our computations, we neglect the widths 
since they are very small compared to the masses considered.
Of course,  one could always tune
$2 m_{\cnone}$ to some hadronic resonance or threshold in order to
drastically increase the cross section and thus reduce the thermal
relic density, but we do not employ such precision tuning.

The squared amplitudes of Eqs.~(\ref{crossx1}) and (\ref{crossx2})
can be used to obtain the thermally averaged annihilation cross
section \cite{falkellis}. Using the notation $s_0=4\mcnone^2$, we have 
\begin{eqnarray}
\langle \sigma v\rangle &=& \frac{\omega(s_0)}{m^2_{\cnone}} - \frac{3}{m_{\cnone}}\bigg[\frac{\omega(s_0)}{m^2_{\cnone}} - 2 \omega^{\prime}(s_0) \bigg]T +\mathcal{O}(T^2) \\ \nonumber
&=& \frac{1}{m^2_{\cnone}} \bigg[1- \frac{3T}{m_{\cnone}} \bigg] \omega(s)\bigg|_{s \rightarrow 4 m^2_{\cnone} + 6 m_{\cnone} T} +\mathcal{O}(T^2),
\end{eqnarray}
where $T$ is the temperature. Keeping terms to zeroth and first order in $T$
should be sufficient for the relic abundance calculation. Writing this as an
expansion in $x=T/m_{\cnone}$, $\langle \sigma v\rangle = a + b x +
\mathcal{O}(x^2)$, we arrive at:
\begin{eqnarray}
\label{sigmava}
a_{\chi \chi \rightarrow A \rightarrow f \bar{f}} &=& \frac{g^4_2 c_f m^2_f \cos^4 \theta_A  \tan^2 \beta}{8 \pi m^2_W} \frac{m^2_{\cnone} \sqrt{1-m^2_f/m^2_{\cnone}}}{(4 m^2_{\cnone}-m^2_A)^2 + m^2_A \Gamma^2_A} \\ \nonumber
\times \bigg[-\epsilon_u (\epsilon_W&-&\epsilon_B \tan \theta_W) \sin \beta + \epsilon_d (\epsilon_W-\epsilon_B \tan \theta_W) \cos \beta  \\ \nonumber
&+& \sqrt{2} \frac{\lambda}{g_2}\epsilon_s (\epsilon_u \sin \beta + \epsilon_d \cos \beta) +\frac{\tan \theta_A}{g_2}\sqrt{2} (\lambda \epsilon_u \epsilon_d -\kappa \epsilon^2_s)  \bigg]^2, \\  \nonumber \\
b_{\chi \chi \rightarrow A \rightarrow f \bar{f}} &\simeq& 0, 
\label{sigmav}
\end{eqnarray}
where $c_f$ is a color factor, equal to 3 for quarks and 1 otherwise.
For this result, we have assumed that the final state fermions are
down-type. If they are instead up-type fermions, the couplings used must be modified as described above. 

We have not written the result for CP-even Higgs exchange because the low velocity term in the expansion is
zero: $a_{\chi \chi \rightarrow H \rightarrow f \bar{f}} = 0$. Although
the $b$-terms can, in principle, contribute to the freeze-out
calculation,
in the computations here such contributions do not have a significant
impact.


The annihilation cross section can now be used to calculate the thermal relic abundance present today.
\begin{equation}
    \label{omegah2}
\Omega_{\cnone} h^2 \approx \frac{10^9}{M_{\rm{Pl}}}\frac{x_{\rm{FO}}}{\sqrt{g_{\star}}} \frac{1}{(a + 3 b/x_{\rm{FO}})},
\end{equation}
where $g_{\star}$ is the number of relativistic degrees of freedom available at freeze-out and $x_{\rm{FO}}$ is given by:
\begin{equation}
    \label{xfo}
x_{\rm{FO}} \approx \ln \bigg(\sqrt{\frac{45}{8}} \frac{m_{\cnone} M_{\rm{Pl}} (a+6b/x_{\rm{FO}}) }{\pi^3  \sqrt{g_{\star} x_{\rm{FO}} }}\bigg).
\end{equation}
For the range of cross sections and masses we are interested in, $x_{\rm{FO}} \approx 20$.

\begin{figure}[tb]
\includegraphics[scale=0.7]{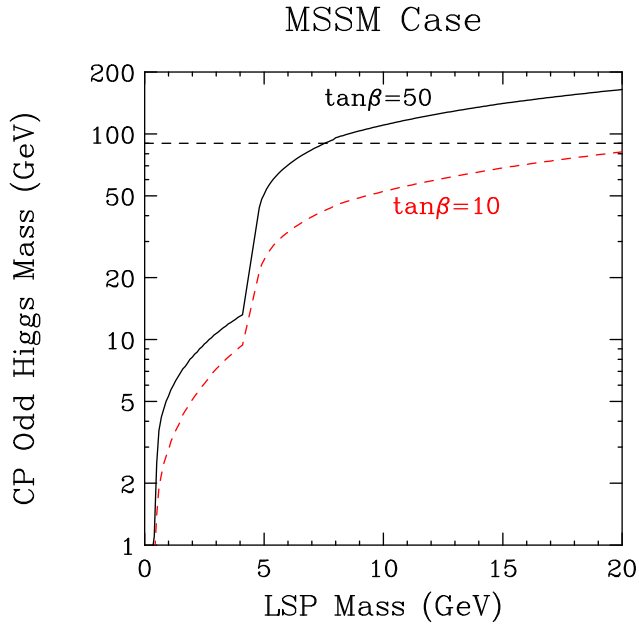}
\caption{
The CP-odd Higgs mass required to obtain the measured relic density
for a light neutralino in the MSSM. Models above the curves produce
more dark matter than in observed. These results are for the case of a
bino-like neutralino with a small higgsino admixture ($\epsilon^2_B =
0.94$, $\epsilon^2_u = 0.06$). Results for two values of $\tan \beta$
(10 and 50) are shown. The horizontal dashed line represents the lower
limit on the CP-odd Higgs mass in the MSSM from collider
constraints. To avoid overproducing dark matter, the neutralino must
be heavier than about 8 (22) GeV for $\tan \beta=50$ (10).}
\label{mssmrelic}
\end{figure}

As a benchmark for comparison, we consider a light bino which annihilates
through the exchange of an MSSM-like CP-odd Higgs ($\cos \theta_A =
1$). The results for this case are shown in Fig.~\ref{mssmrelic}. In
this figure, the thermal relic density of LSP neutralinos exceeds the
measured value for CP-odd Higgses above the solid and dashed curves,
for values of $\tan \beta$ of 50 and 10, respectively. Shown as a
horizontal dashed line is the lower limit on the the MSSM CP-odd Higgs
mass from collider constraints. This figure demonstrates that even in
the case of very large $\tan \beta$, the lightest neutralino must be
heavier than about 7 GeV. For moderate values of $\tan \beta$, the
neutralino must be heavier than about 20 GeV.

\begin{figure}[tb]
\includegraphics[scale=0.6]{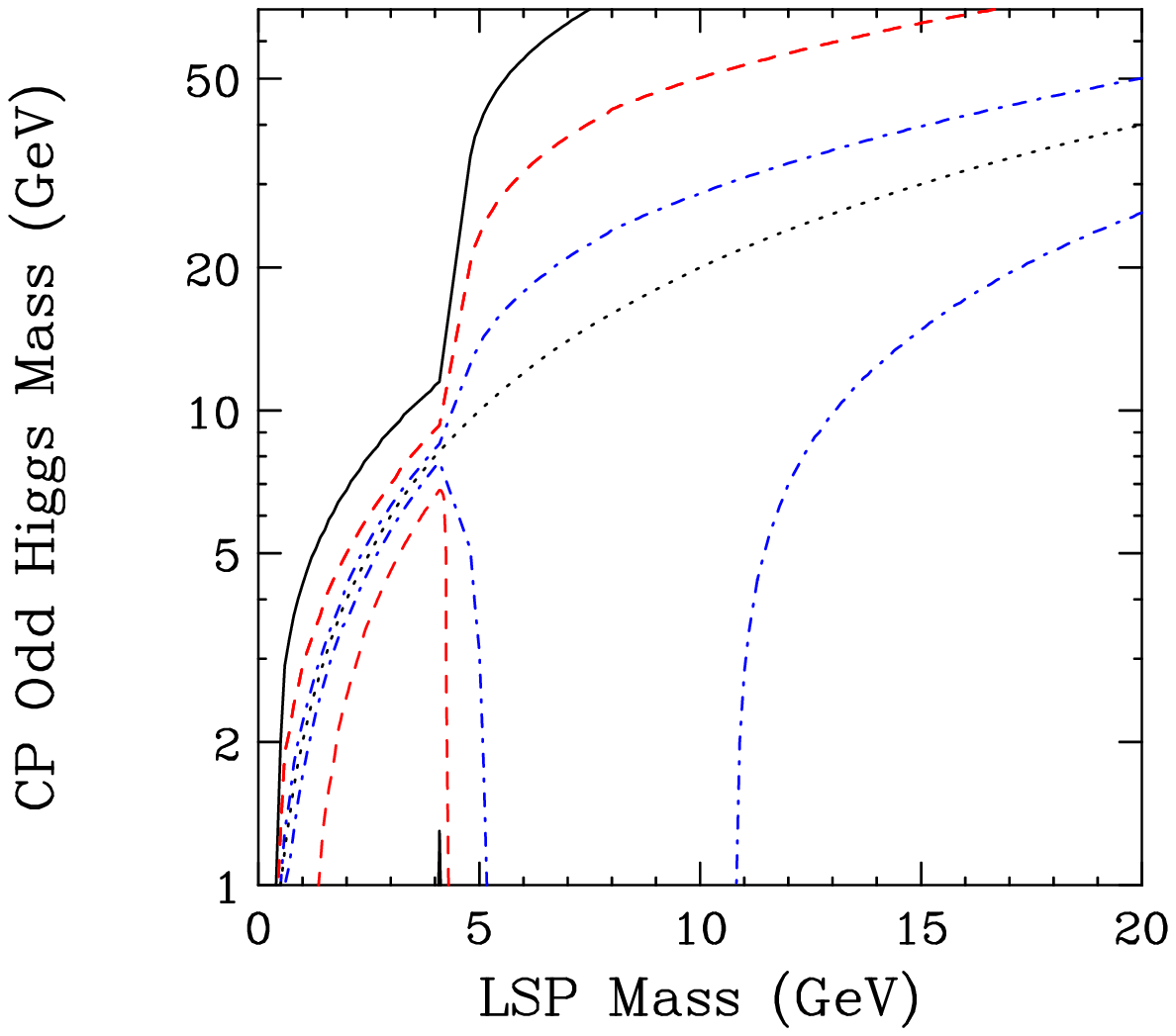}
\includegraphics[scale=0.6]{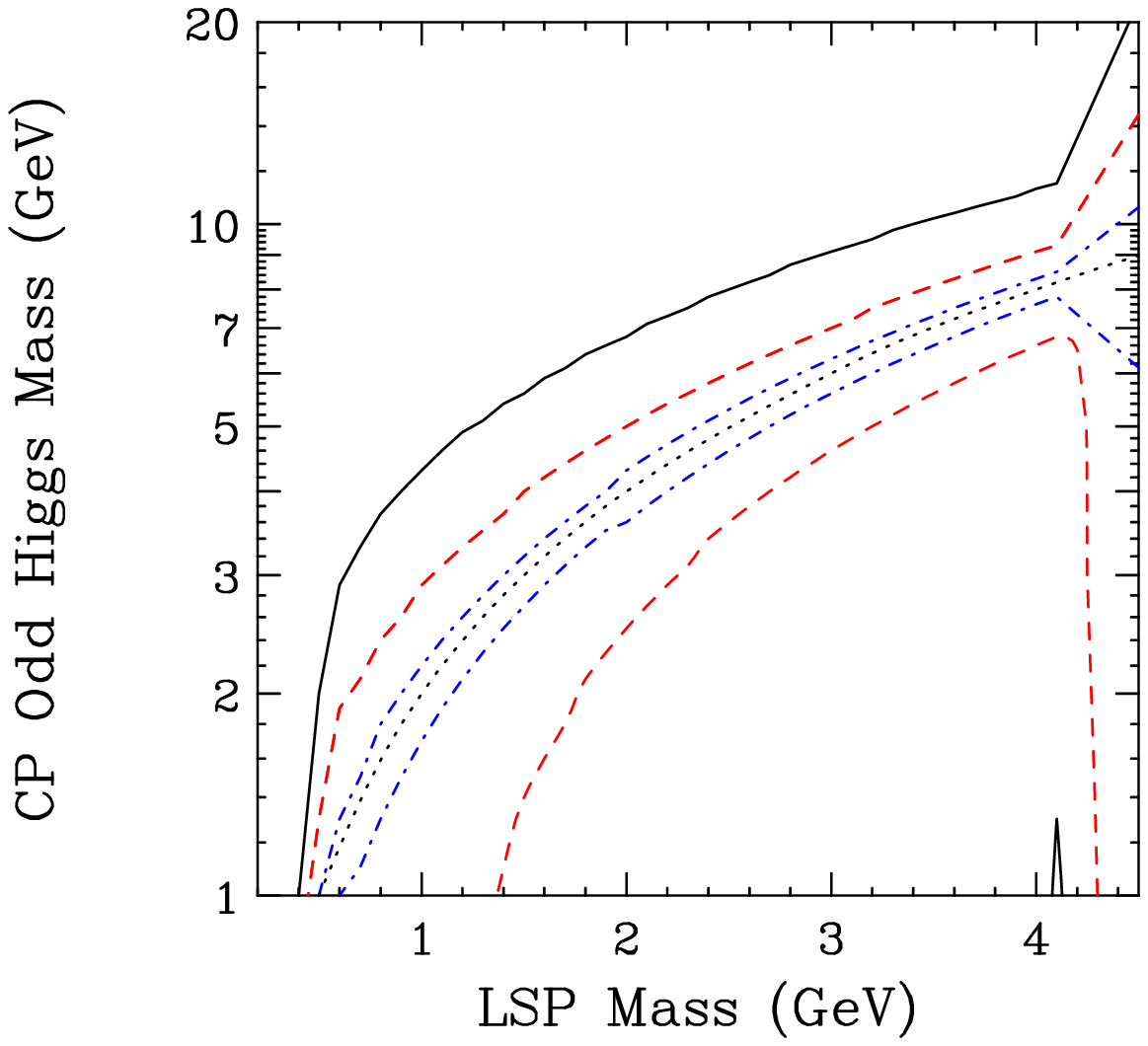}
\caption{
We display contours in $m_{A_1}$ -- $\mcnone$
parameter space for which Eq.~(\ref{omegah2}) yields $\Omega h^2=0.1$.
Points above or below each pair of curves produce more dark matter than
is observed; inside each set of curves less dark matter is produced than
is observed.  These results are for a bino-like neutralino with a small
higgsino admixture ($\epsilon^2_B = 0.94$, $\epsilon^2_u = 0.06$). Three
values of $\tan \beta$ (50, 15 and 3) have been used, shown as solid
black, dashed red, and dot-dashed blue lines, respectively. 
The dotted line is
the contour corresponding to $2 m_{\cnone}=m_A$. For each set of lines,
we have set $\cos^2 \theta_A =0.6$. The $\tan\beta =50$ case is highly
constrained for very light neutralinos, and is primarily shown for
comparison with the MSSM case.}
\label{nmssmrelic}
\end{figure}

In Fig.~\ref{nmssmrelic}, we show how this conclusion is modified
within the framework of the NMSSM. Here, we have considered a CP-odd
Higgs which is a mixture of MSSM-like and singlet components 
specified by $\cos^2
\theta_A = 0.6$ and a neutralino with composition specified
by $\eps_B^2=0.94$ and $\eps_u^2=0.06$. These specific
values are representative of those that can be achieved for
various NMSSM parameter choices satisfying all constraints.
For each pair of contours (solid black, dashed red,
and dot-dashed blue), the region between the lines is the space in
which the neutralino's relic density does not exceed the measured
density. The solid black, dashed red, and dot-dashed blue lines
correspond to $\tan \beta$=50, 15 and 3, respectively. Also shown as a
dotted line is the contour corresponding to the resonance condition,
$2 m_{\cnone}=m_{A}$.

For the $\tan \beta$=50 or 15 cases, neutralino dark matter can avoid
being overproduced for any $A_1$ mass below $\sim 20-60$ GeV, as long
as $m_{\cnone} > m_b$. For smaller values of $\tan \beta$, a lower
limit on $m_{A_1}$ can apply as well.

For neutralinos lighter than the mass of the $b$-quark, annihilation is
generally less efficient. This region is shown in detail in the right
frame of Fig.~\ref{nmssmrelic}. In this funnel region, annihilations
to $c \bar{c}$, $\tau^+ \tau^-$ and $s \bar{s}$ all contribute
significantly. Despite the much smaller mass of the strange quark, its
couplings are enhanced by a factor proportional to $\tan \beta$ (as
with bottom quarks) and thus can play an important role in this mass
range. In this mass range, constraints from Upsilon and $J/\psi$
decays can be very important, often requiring fairly small values of
$\cos \theta_A$.

For annihilations to light quarks, $c \bar{c}$, $s \bar{s}$, etc., the
Higgs couplings to various meson final states should be considered,
which include effective Higgs-gluon couplings induced through quark
loops. In our calculations here, we have used the conservative
approximation of the Higgs-quark-quark couplings alone, even for these
light quarks, but with kinematic thresholds set by the mass of the
lightest meson containing a given type of quark, rather than the quark
mass itself. This corresponds to thresholds of 9.4 GeV, 1.87 GeV, 498
MeV and 135 MeV for bottom, charm, strange and down quarks,
respectively. A more detailed treatment, which we will not undertake
here, would include the proper meson form factors as well as allowing
for the possibility of virtual meson states.

Thus far, we have focused on the case of a bino-like LSP. If the LSP
is mostly singlino, it is also possible to generate the observed relic
abundance in the NMSSM. A number of features differ for the
singlino-like case in contrast to a bino-like LSP, however. First, the
ratio $\mcnone/m_{A_1}$ cannot be arbitrarily small. The relationship
between these two masses was shown for singlino-like LSPs in
Fig.~\ref{l_k_chiovera}. As discussed earlier, and shown in
this figure, an LSP mass
that is chosen to be precisely at the Higgs resonance, $m_{A_1}\simeq
2\mcnone$, is not possible for this case: $m_{A_1}$ is always less
than $2\mcnone$ by a significant amount.

Second, in models with a singlino-like LSP, the $A_1$ is generally
also singlet-like and the product of $\tan^2 \beta$ and
$\cos^4 \theta_A$ is typically very small. This limits the ability of a
singlino-like LSP to generate the observed relic abundance.  The last term in
Eq.~(\ref{sigmava}) introduces an additional $\tan^2 \theta_A$ dependence,
however, which effectively reduces the impact of $\cos \theta_A$ on the
annihilation cross section from four powers to two. But, this last
term is  suppressed when the singlet fraction $\eps_s$ is large and
$\eps_u,\eps_d$ are small 
by the factor of $\kappa$ (which is small for a singlino) that
multiplies
$\eps_s^2$. Alternatively, the second
to last term in Eq.~(\ref{sigmava}) can also be of importance.  
Overall, the inability to compensate the smallness of
the coefficients in Eq.~(\ref{sigmava}) 
by being nearly on-pole implies that 
annihilation is too inefficient for an LSP that is more than 80\%
singlino. 

In the following section, we give sample cases
 for which $m_{A_1}$ and $\mcnone$ are light and  
$\Omega h^2\sim 0.11$.  These are representative of the many different
types of scenarios that are possible and include a case in which
the $\cnone$ is largely singlino.

\section{Sample Model Points}

In this section we present specific sample model points of the type we
propose.  These points are obtained using NMHDECAY 1.1.~\cite{Ellwanger:2004xm}

The first has a singlet-like $H_1$, which would have escaped detection
at LEP due to this singlet nature.  
In addition, the mass of the more SM-like $H_2$ is beyond the LEP reach.
It also has a sizable $\br(\Upsilon
\rightarrow \gamma+A_1)$ which could be discovered by a re-analysis of
existing CLEO data.

\begin{table}[ht]
\begin{tabular}[b]{c|c|c|c|c|c|c|c}
    $\lambda$& $\kappa$     & $\tan \beta$ & $\mu$ & $A_\lambda$ & $A_\kappa$ & $M_1$ & $M_2$ \\
    0.436736 & -0.049955    & 1.79644      & -187.931 & -458.302 & -40.4478 & 1.92375 & 390.053 \\
    \hline
    $M_{A_1}$&  $\cos \theta_A$ \\
    7.17307  &  -0.193618       \\
    \hline
    $M_{H_1}$& $\xi_u$ & $\xi_d$ & $\xi_s$ \\
    73.8217  & 0.1127  & -0.0277 & 0.9932  \\
    \hline
    $M_{\cnone}$ & $\epsilon_{{B}}$ & $\epsilon_{{W}}$ & $\epsilon_u$ & $\epsilon_d$ & $\epsilon_s$ \\
    3.49603      & -0.781466              & -0.00594669            & 0.11476      & 0.26493      & 0.553099 \\
    \hline
    $\br(\Upsilon \rightarrow \gamma + A_1)$ & $\langle \sigma v \rangle$ & $\Omega h^2$ \\
    8.12331e-06 & 4.55841e-26 $cm^3/s$       & 0.107689 \\
\end{tabular}
\caption{Sample model point \#1.}
\end{table}

The second point has an MSSM-like $H_1$, but due to the presence of the light
$A_1$ and the large $\lambda$ coupling, this MSSM-like $H_1$ decays
dominantly to a pair of  $A_1$'s  [$BR(H_1\rightarrow A_1 A_1)=99.6\%$ for this
point]. Such an $H_1$ would not be easily detected at the LHC.

\begin{table}[ht]
\begin{tabular}[b]{c|c|c|c|c|c|c|c}
    $\lambda$& $\kappa$     & $\tan \beta$ & $\mu$ & $A_\lambda$ & $A_\kappa$ & $M_1$ & $M_2$ \\
    0.224982 & -0.47912    & 7.58731      & -174.624 & -421.908 & -30.6106 & 21.0909 & 984.116 \\
    \hline
    $M_{A_1}$&  $\cos \theta_A$ \\
    46.6325  &  -0.570716       \\
    \hline
    $M_{H_1}$& $\xi_u$ & $\xi_d$ & $\xi_s$ \\
    117.72  & 0.9823  & 0.1848 & 0.0316  \\
    \hline
    $M_{\cnone}$ & $\epsilon_{{B}}$ & $\epsilon_{{W}}$ & $\epsilon_u$ & $\epsilon_d$ & $\epsilon_s$ \\
    22.37      & -0.9715              & -0.0024            & 0.0020      & 0.2366      & 0.0128 \\
    \hline
    $\br(\Upsilon \rightarrow \gamma + A_1)$ & $\langle \sigma v \rangle$ & $\Omega h^2$ \\
    0       & 2.17478e-25 $cm^3/s$       & 0.108649
\end{tabular}
\caption{Sample model point \#2.}
\end{table}

The third point has a singlino-like $\tilde{\chi^0_1}$
as well as a singlet-like $H_1$.  As for point
\#1, this point 
has a $\br(\Upsilon \rightarrow \gamma+ A_1)$ that might be excluded
by an appropriate re-analysis of existing data.

\begin{table}[ht]
\begin{tabular}[b]{c|c|c|c|c|c|c|c}
    $\lambda$& $\kappa$     & $\tan \beta$ & $\mu$ & $A_\lambda$ & $A_\kappa$ & $M_1$ & $M_2$ \\
    0.415867 & -0.029989   & 1.78874      & -175.622 & -455.387 & -39.671 & 7.1098 & 289.115 \\
    \hline
    $M_{A_1}$&  $\cos \theta_A$ \\
    8.35008  &  -0.187349       \\
    \hline
    $M_{H_1}$& $\xi_u$ & $\xi_d$ & $\xi_s$ \\
    63.3851 & -0.1412 & -0.1810 & 0.9733  \\
    \hline
    $M_{\cnone}$ & $\epsilon_{{B}}$ & $\epsilon_{{W}}$ & $\epsilon_u$ & $\epsilon_d$ & $\epsilon_s$ \\
    -3.98   & -0.3697              & -0.0262            & 0.2524      & 0.2560      & 0.8564 \\
    \hline
    $\br(\Upsilon \rightarrow \gamma + A_1)$ & $\langle \sigma v \rangle$ & $\Omega h^2$ \\
    3.96e-6 & 4.12241e-26 $cm^3/s$       & 0.119239
\end{tabular}
\caption{Sample model point \#3.}
\end{table}

\section{Elastic Scattering of Light Neutralinos}

The spin-independent elastic scattering cross section of a light
neutralino with nuclei is generally dominated by the $t$-channel exchange
of a CP-even Higgs boson. The cross
section for this process is approximately given by:
\begin{equation}
\sigma_{\rm{elastic}} \approx \sum_{H} 
\frac{1}{\pi m_H^4 } 
\left(  \frac{m_p
m_{\cnone}}{m_p +m_{\cnone}}  \right)^2 C_{\cnone \cnone H}^2 \left( \sum_q C_{q\anti q H}\langle N|q \bar q|N
\rangle \right)^2
\label{sigelastic}
\end{equation}
where the first sum is over the CP-even Higgs states of the NMSSM
and $m_H$ are their masses. The second sum is over the quark
types and $<N|q \bar{q}|N>$ are the matrix elements over the atomic nuclear
state. Of course, one must be careful to use the correct form
of $C_{q\anti qH}$ which differs for up-type quarks versus down-type quarks.
In the sum over quark species, the strange quark
contribution dominates with $m_s <N|s \bar{s}|N> \approx 0.2$ GeV. For
a bino-like LSP and any one $H$, Eq.~(\ref{sigelastic}) reduces to
\bea
\sigma^{\rm{bino}}_{\rm{elastic}} &\sim &\frac{8 G_F^2 m_Z^2}{\pi m_H^4}
\bigg(\frac{m_p m_{\cnone}}{m_p+ m_{\cnone}}\bigg)^2  \, \epsilon^2_B
\sin^2\theta_W \, (\eps_d\xi_u-\eps_u\xi_d)^2 \, \times
\nonumber\cr
&&\quad
\bigg(
\sum_{q=d,s,b} \frac{m_q\xi_d}{\cos \beta} <N|q \bar{q}|N> +\sum_{q=u,c}
\frac{m_q\xi_u}{\sin\beta} <N|q \bar{q}|N>\bigg)^2.
\eea
If the LSP is singlino-like, on the other hand, the appropriate
approximation is
\bea
\sigma^{\rm{singlino}}_{\rm{elastic}} &\sim& \frac{8 G_F^2 m_Z^2}{\pi m_H^4}
\bigg(\frac{m_p m_{\cnone}}{m_p+ m_{\cnone}}\bigg)^2 \,  \frac{ 2 \lambda^2 \,
\epsilon^2_s  \, \cos^2 \theta_W}{g^2_2} \, (\eps_d\xi_d+\eps_u\xi_u)^2 \,\times 
\nonumber\cr
&&\quad
\bigg(
\sum_{q=d,s,b} \frac{m_q\xi_d}{\cos \beta} <N|q \bar{q}|N> +\sum_{q=u,c}
\frac{m_q\xi_u}{\sin\beta} <N|q \bar{q}|N>\bigg)^2.
\eea
where, in $C_{\cnone\cnone H}$, we have dropped the term
containing $\kappa$ since it
is expected to be very small and the term
proportional to $\epsilon_u \,\epsilon_d$ which is also likely to be
very small.

In assessing the implications of the above, it is useful to
note that LEP limits on a Higgs boson with $m_H<120\gev$ generally imply 
\begin{equation}
\xi_{u,d} \lsim \left(\frac{m_H}{120 \rm GeV}\right)^{3/2} + 0.1,
\label{xilim}
\end{equation}
and for a light $\cnone$ LEP limits
on invisible $Z$ decays roughly imply $\epsilon_{u,d} < 0.06$.

The claim of a positive WIMP detection made by the DAMA collaboration is
not consistent with the limits placed by CDMS and others for a WIMP in
the mass range normally considered (above a few tens of GeV). Very light
WIMPs, however, scatter more efficiently with light target nuclei than
with heavier nuclei, which can complicate this picture. For a WIMP with
a mass between about 6 and 9 GeV, it has been shown that the DAMA
results can be reconciled with the limits of CDMS and other
experiments~\cite{gondolo}.\footnote{If a tidal stream of dark matter is
present in the local halo, WIMP masses over a somewhat wider range can
reconcile DAMA with CDMS as well.} This is made possible by the
relatively light sodium (A=23.0) component of the DAMA experiment
compared to germanium (A=72.6) and silicon (A=28.1) of CDMS. 

To produce the rate observed by DAMA, a light WIMP would need an elastic
scattering cross section of 7$\times 10^{-40}$ cm$^2$ to 2$\times 10^{-39}$
cm$^2$ ($0.7-2$ fb).  
For the case of a bino-like or singlino-like neutralino capable of
resolving the DAMA discrepancy, the scale of this cross section is:
\begin{equation}
    \label{elastic_estimate}
    \sigma_{\rm{elastic}} \lsim 1.4\times 10^{-42} { cm^2} \left(\frac{120~
    {\rm GeV}}{m_H}\right)^4\left(\left(\frac{m_H}{120 ~{\rm
    GeV}}\right)^{3/2}+0.1\right)^2 \left(\frac{\tan\beta}{50}\right)^2 F_\lambda
\end{equation}
assuming $m_{\cnone} > m_p$ and $\tan \beta > 1$, using the
$\xi_{u,d}$ limit of Eq.~(\ref{xilim}) and adopting $\eps_{u,d}\sim 0.06$.
One has
$F_{\lambda} = 1$ for the bino-like case and $F_{\lambda} = 2 \lambda^2
/(g^2_2 \tan^2 \theta_W) \approx 0.67 \times (\lambda/0.2)^2$ for the
singlino-like case. For $\tan \beta = 50$,
$\lambda = 0.2$ and a Higgs mass of 120 GeV, we estimate a
neutralino-proton elastic scattering cross section on the order of
$4\times 10^{-42}$ cm$^2$ ($4 \times 10^{-3}$ fb) for either a bino-like or a singlino-like
LSP. This value may be of interest to direct detection searches such as CDMS,
DAMA, Edelweiss, ZEPLIN and CRESST.  To account for the DAMA data, the cross
section would have to be enhanced by a local over-density of dark matter~\cite{gondolo}.

The cross section in Eq.~(\ref{elastic_estimate}) is
small unless $\tan \beta$ is quite large, in which case
the scenario will run into
difficulty with LEP limits unless $\cos \theta_A$ is quite small.  To
explain the DAMA result, we can instead require $m_H$ to be small.
For instance, with $\mcnone=6$ GeV, $m_H=3$ GeV, and $\tan \beta=10$,
the DAMA result can be reproduced with $\sigma_{\rm elastic} \sim
4\times 10^{-39}
{\rm cm}^2$ ($\sim 4$ fb), {\it without} requiring a dark matter wind through
our solar system.  It would not be unusual for a mostly-singlet $H_1$ to be this
light if $\lambda$ is small.  In this case the singlet decouples from the MSSM
and the whole singlet supermultiplet is light.  

For a detailed study of direct detection prospects for heavier
neutralinos in the NMSSM, see Refs.~\cite{nmssmdm,Belanger:2005kh}.  We
find consistency with their results concerning annihilation through $H$
and $A$ resonances.

\section{Extremely Light Neutralinos and the Observation of 511 keV
Emission From the Galactic Bulge}

If the LSP's mass is even smaller, below $\sim1$ GeV, it may still be
possible to generate the observed relic density. In this mass range, in addition to annihilations to
strange quarks ($K^{\pm}$, $K^0$), final state fermions can include muons and even lighter quarks ($\pi^{\pm}$, $\pi^0$).

There is a $\cnone$ mass range in which neutralinos will annihilate
mostly to muon pairs.  This range is $m_\mu < m_{\cnone} < m_{\pi^+} +
m_{\pi^0}/2$, or $106$ MeV $< m_{\cnone} < 207$ MeV. (The upper limit
will be explained shortly.) This range of
parameter space is of special interest within the context of the 511 keV
emission observed from the galactic bulge by the INTEGRAL/SPI
experiment. Muons produced in neutralino annihilations will quickly
decay, generating electrons with energies of $\sim m_{\cnone}/3$, which
may be sufficiently small for them to come to rest in the galactic bulge
before annihilating.

The upper limit above derives from the fact that the $\cnone\cnone$
annihilations should not create many $\pi^0$'s. In this way, we avoid
gamma ray constraints from EGRET.  If we assume that the annihilation
mediator is the CP-odd $A_1$, 
$\cnone\cnone\to A_1\to\mbox{pions}$ is only possible if $2\mcnone\gsim
2m_{\pi^+}+m_{\pi^0}$ since the lowest threshold channel
is to three pions: $\cnone\cnone \rightarrow A_1
  \rightarrow \pi^+ \pi^- \pi^0$.
Also note that by generating
positrons through muon decays rather than directly allows gamma ray
constraints from final state radiation~\cite{Beacom:2004pe} to be
easily evaded.

It has been shown that a $\sim$100 MeV dark matter particle annihilating
through an $a$-term (low velocity) cross section can simultaneously yield
the measured relic density and generate the number of positrons needed
to accommodate the INTEGRAL/SPI data \cite{511annihilate}. These are
precisely the features of a 106-207 MeV neutralino combined with
the presence of a 100 MeV--1 GeV CP-odd Higgs.

The main difficulty with this scenario comes from the constraints on
Upsilon decays, which we discussed in Sec.~\ref{upsconstraints}. To
evade the CLEO 
limit~\cite{ups1stogammainvis} of $BR(\Upsilon \rightarrow \gamma A_1)
< 2\times 10^{-5}$ in this mass region, we must require $\cos^2\theta_A
\tan^2\beta < 0.13$ [see Eq.~(\ref{vaxiondecay})]. 
Given these constraints, and considering a 
bino-like neutralino with a 6\% higgsino admixture and
$m_{\cnone}=150$ MeV, the 
annihilation cross section needed to avoid overproducing dark matter
can only be 
attained for a fairly narrow range of $m_{A_1} \approx 2  m_{\cnone}
\pm 10$ MeV. 
This scenario, although not particularly attractive due to this
requirement, 
does demonstrate that it is possible to generate the INTEGRAL signal with
neutralinos in the NMSSM.  This can be confirmed or ruled out by
improving the 
limit on $BR(\Upsilon \rightarrow \gamma A_1)$ where the $A_1$ is not
observed 
or where the $A_1$ decays to a muon pair.  In the latter case, the
$A_1$ may 
have a significant displaced vertex of a few cm, especially for small
$\tan 
\beta$ and $m_{A_1} < 2 m_{\cnone}$~\cite{lighta}.

An $A_1$ this light (300 MeV) is too light to be technically natural, however.
Radiative corrections pull up its mass and a cancellation between different
orders in perturbation theory is required for $A_1$ to be this light.  While we
have found parameter points capable of yielding the INTEGRAL
signal, we find that they are
not stable in the sense that if any of the Higgs-sector parameters are adjusted
by a very small amount, the $A_1$ is pulled up in mass to $\mathcal{O}(10\gev)$.
From our numeric analysis, $m_{A_1}$ as small as a few GeV is technically
natural.


\section{Conclusions}

In this article, we have studied the possibility of light neutralinos
(100 MeV to 20 GeV) being present within the
Next-to-Minimal-Supersymmetric-Standard-Model (NMSSM)
without conflicting with constraints on the dark matter of the universe.
We find that light CP-odd Higgs bosons with a substantial non-singlet
component, which appear naturally within this framework,
can provide an efficient annihilation channel for light, bino or singlino-like
neutralinos. This channel makes it possible for very light neutralinos
to generate the observed dark matter abundance, unlike in the case of
neutralinos in the MSSM.

Within this model, we have discussed the implications of light
neutralinos for direct detection and find that the NMSSM can naturally
provide neutralinos in the mass range (6-9 GeV) as required to reconcile
the DAMA claim of discovery with the limits placed by CDMS and other
experiments. We have also explored the possibility that the 511 keV
emission observed from the galactic bulge by INTEGRAL/SPI could be
generated through neutralino annihilations into muon pairs. This
scenario appears possible for a very light (106-207 MeV) neutralino
and for a light CP-odd Higgs boson with mass close to twice
the neutralino mass, provided $\tan \beta$ is not large.

This kind of scenario containing a light neutralino and/or light
axion-like particles represents a challenge for the LHC and ILC, and is
deserving of further analysis.  We note that only the ILC will be able
to study the properties of the $\cnone$ and $A_1$ adequately to verify
that they are consistent with the observed dark matter density.


\vskip 0.5in
\vbox{
\noindent{ {\bf Acknowledgments} } \\
\noindent
This work was supported in part by DOE grant
DOE--EY--76--02--3071, the 
Davis Institute for High Energy Physics, and the U.C. Davis Dean's office. DH is supported by the Leverhulme trust.  JFG thanks the Aspen Center for Physics for support during the completion
of this work.}

\pagebreak
\renewcommand{\theequation}{A.\arabic{equation}}
\renewcommand{\thesection}{A-\arabic{section}}
\setcounter{equation}{0}  
\setcounter{section}{0}
\pagebreak

\end{document}